\documentclass[12pt]{article}
\begin{document}
\title{Tail probabilities for short-term returns on stocks}
\author{Henrik O. Rasmussen$^1$ \ \& P. Wilmott$^2$}
\date{%
$^1$ OCIAM, Mathematical Institute,  \\ %
University of Oxford,  24 -- 29 St. Giles, \\ %
OX1 3LB Oxford, UK \\ %
henrik@hibrium.com \\ [2ex]%
$^2$ Wilmott Associates \\ [2ex]%
DOI: 10.13140/RG.2.2.18816.28165 \\ [2ex]%
This version: 2 March 2003; Updated 20 March 2019}
\begin{titlepage}
\maketitle\begin{abstract}
We consider the tail probabilities of stock returns for a general class of stochastic volatility models. In these models, the stochastic differential equation for volatility is autonomous, time-homogeneous and dependent on only a finite number of dimensional parameters. Three bounds on the high-volatility limits of the drift and diffusion coefficients of volatility ensure that volatility is mean-reverting, has long memory and is as volatile as the stock price. Dimensional analysis then provides leading-order approximations to the drift and diffusion coefficients of volatility for the high-volatility limit. Thereby, using the Kolmogorov forward equation for the transition probability of volatility, we find that the tail probability for short-term returns falls off like an inverse cubic. Our analysis then provides a possible explanation for the inverse cubic fall off that Gopikrishnan {\it et al.} (1998) report for returns over $5$ -- $120$ minutes intervals. We find, moreover, that the tail probability scales like the length of the interval, over which the return is measured, to the power $3/2 $. There do not seem to be any empirical results in the literature with which to compare this last prediction.
\end{abstract}
\end{titlepage}
\section{Introduction}
It has long been known \cite{ma,fa} that the tails of the distribution of returns on stocks are much fatter than they would be if these returns were normal distributed, but it is only recently that these tails have been accurately resolved. In particular, Ref.\cite{go1} reports that 
\begin{equation} \label{inv}  \bar P(x) \propto x^{-3},
\end{equation}
for $x$ ranging over approximately two decades, where $\bar P$ is the tail probability \[  \bar P(x) = {\rm Prob}(\vert X \vert \geq x),\]
for the logarithmic return
\[  X = \ln \biggl [ \, {{S(t+\Delta t)} \over {S(t)}} \,       \biggr ]
\]on a stock price $S(t)$. 
The authors verify this `inverse cubic' fall off for $\Delta t$ from $5$ to $120$ minutes. (In Ref. \cite{go1}, a mean value is subtracted from $X$, but this mean value is negligible when compared with the fluctuating part, when $\Delta t$ is this small \cite{hull}). The underlying data set is all prices quoted over a two--year period for the $1000$ largest companies listed on the three major US stock markets. 
\par
The purpose of the present paper is to suggest an explanation for the result in Eq. \ref{inv}. We shall consider only a single stock. Instead of averaging over stocks, as in Ref. \cite{go1}, we average over time. I.e., we consider the stationary probability density of the return for a single stock. Subject to a weak ergodic hypothesis, the two types of averaging produce identical results for the rate at which the tail probability $\bar P(x)$ falls off for large $x$. (See also \cite{bou}). 
\par
We suppose that the stock price process $S$ and its volatility $\sigma $ satisfy stochastic differential equations of the type \cite{me,du,wi},
\begin{eqnarray} \label{bs}   
dS/S  &  = &  \mu \, dt + \sigma \, dW_1 \\   d\sigma &  = & \alpha \, dt + \beta \, dW_2,    
\label{vol}\end{eqnarray}
where $W_1$ and $W_2$ are standard Wiener processes, possibly correlated. In the following, we use dimensional analysis to obtain a stochastic volatility model for the high--volatility limit. The starting point is a generalisation of some of the stochastic volatility models that have been considered previously in the literature. In these models, the drift 
$\alpha $ and diffusivity $\beta $ are functions of only volatility $\sigma$ and a finite number of parameters $\{ r_0,\ldots, r_N \}$, 
\begin{eqnarray} \label{g1}   
\alpha & = & \alpha(\sigma, r_0, \ldots, r_N) \\    
\beta  & = & \beta(\sigma,r_0, \ldots, r_N). 
\label{g2}\end{eqnarray}
Since the dimensions of $\alpha $, $\beta $ and $\sigma $ are all powers of time, there is no loss of generality in taking the dimensions of the parameters $\{ r_0,\ldots,r_N \}$ to be inverse time (as shown below). Each parameter then represents as a characteristic time scale for the volatility process. 
\par
To simplify the argument, however, we assume in most of the following that there is only one dimensional parameter $r_0$. If we require volatility processes to be mean reverting, then it can be shown (by a dimensional argument) that $r_0$ must represent the time scale for mean reversion. Since the observed rates of mean reversion are low, we 
require $\alpha $ and $\beta $ to have well--defined limits as the parameter $r_0$ tends to zero. The easiest 
way to ensure 
this is to assume that $\alpha $ and $\beta $ are differentiable at $r_0=0$. To obtain a model for the high--volatility 
limit, we make use of one more stylised fact about volatility. Namely, that the volatility process itself is at least as 
volatile as the stock price process $S$. (These stylised facts about volatility are discussed in more detail below). Remarkably, these conditions alone provide the high--volatility asymptotic behaviour of $\beta $ and also 
put a bound on the rate of growth of $\alpha $ in the same limit. 
\par
To estimate the tail probability $\bar P$ of the logarithmic return $X$, we then note that 
\begin{equation} \label{ap}   
X =_d \sigma \, \bigl [ \, W_1(t+\Delta t) - W_1(t)        \, \bigr ] + O(\Delta t),
\end{equation}where the subscript $d$ denotes distribution. Since $\sigma $ and the Wiener processare independent random variables, this last identity allows us to express the pdf of $X$ in terms of the pdfs of $\sigma $ and a $N(0,\Delta t)$ distributed random variable. To determine the $p(\sigma,t)$ of $\sigma $, we insert the previously obtained estimates of $\alpha $ and $\beta $ in the Kolmogorov forward equation,
\[  
{{\partial p} \over {\partial t}} + {\partial \over {\partial\sigma}} \biggl [ \, \alpha \, p \, \biggr ] = {1 \over 2} \, {{\partial^2} \over {\partial\sigma^2}} \biggl [ \, \beta^2 \, p \, \biggr ].
\]
\par
The outline is as follows. In section II, we show that the parameters may be taken, without loss of generality, to have the same dimension as the interest rate, namely the inverse of time. In section III, we discuss the conditions ensuring agreement with stylized facts about volatility. In section IV, we argue that the conditions for asymptotic analysis are satisfied in practice. In section V, we present dimensional analysis of $\alpha $ and $\beta $. In section VI, these results provide asymptotic approximations to the high-volatility limits of $\alpha $ and $\beta$. Using these approximations, we derive the stationary probability density of volatility in section VII. In section VIII, we finally derive the tail probability ${\rm P}(x)$. 

\section{The parameters} 
The argument in this paper is based on dimensional analysis of the coefficients $\alpha$ and $\beta $ appearing in Eq. \ref{vol}. The formal basis for dimensional analysis is the so-called $\Pi$-theorem \cite{bi,ba}, which states that any non-dimensional function of a finite number of dimensional variables depends only on finitely many non-dimensional combinations of these variables. This theorem simplifies the following analysis and also shows that there is no essential difference, in the high--volatility limit, between the cases where $\alpha $ and $\beta $ depend on only one parameter and where they depend on several parameters.
\par
As a first application of the $\Pi$-theorem, let us show that there is no loss of generality in assuming that the parameters $\{r_0,\ldots,r_N\}$ have the dimension inverse time. For the sake of simplicity, we suppose that there are only two parameters, $r_0$ and $r_1$. After non-dimensionalising $\alpha$ and $\beta$ by means of $\sigma$ (see section 5), it follows from the $\Pi$-theorem that the resulting functions depend only on non-dimensional combinations of $\sigma $,$r_0$ and $r_1$. Since $\sigma$ has the dimension of an inverse square root of time, some combination of $r_0$ and $r_1$ must yield a quantity with the dimension square-root time. If this combination is 
\begin{equation}  r_0 \, r_1^{\gamma} 
\end{equation}
then 
\begin{equation} 
r_0' = r_0^{-2} \, r_1^{-2 \gamma} 
\end{equation}has the dimension of inverse time. If $r_0$ and $r_1$ originally had dimensions that were powers of time, one more quantity with the dimension inverse time can be found. Otherwise, $r_0'$ is the only such quantity. The $\Pi$-theorem now ensures that any non--dimensional function of $(\sigma,r_0,r_1)$ can be written as a non--dimensional function of only $(\sigma,r'_0)$. A similar argument can be used when there are more than two parameters. 
\par
In addition to $\{r_0, \ldots, r_N\}$, there may be a much larger parameter with the dimension of inverse time, associated with the smallest possible time in which the market can react to news. In the foreign-exchange market, this time-scale is of the order of one minute, and a similar magnitude seems plausible for stocks. In the following, we consider this reaction time simply as the `tick-time' used when passing from a discrete to a continuous model of trading \cite{me}, so the corresponding rate-of-change can be ignored here.
\par
Since we consider the case where $\sigma^2$ is much larger than any parameter, it is not necessary to determine these parameters. But to verify that the conditions for asymptotic analysis are satisfied, we need an order-of-magnitude estimate of the largest parameter. Such an estimate is provided in section 4 for the case where only one parameter $r_0$ appears in the model.

\section{Stylized facts about volatility}
In this section, we discuss three stylized facts with strong empirical support. The need to agree with these stylized facts motivates the following assumptions on $\alpha $ and $\beta $,
\begin{eqnarray} \label{sup}      
\alpha (\sigma,r_0, \ldots, r_N) & \leq &  0 \\    
\label{pers}   
\lim_{\sigma \to \infty} \vert \alpha(\sigma ) \vert \, \sigma^{-3} & = & 0 \\    
\label{vvol}   \liminf_{\sigma \to \infty} \beta \, \sigma^{-2} & > & 0,
\end{eqnarray}
where Eq. \ref{sup} should hold for $\sigma \geq \sigma_{\min}$, for some choice of $\sigma_{\min}$ proportional to the root-mean-square volatility.
\vskip 0.25in
\begin{center}    
{\large \bf Mean reversion}
\end{center}
This feature of volatility is well documented \cite{to}. It refers to the fact that volatility tends to decrease when far above historical mean values,
\begin{equation} 
E\bigl \{ \sigma_s \vert \, {\rm F}_t \bigr \} \> \leq \>  \sigma_t, \qquad s \geq t,
\end{equation}
where ${\rm F}_t$ denotes the sigma-algebra at time $t$ associated with the natural filtration $\{ {\rm F}_t \}$ for the process $\sigma_t$. The left side is then the expectation conditional on information available at time $t$. Since
\begin{equation} \label{drift}   
E\bigl \{ \sigma_s \vert {\rm F}_t \bigr \} = \sigma_t + \int_t^s     E\bigl \{ \alpha(\sigma_{\tau}) \vert \, {\rm F}_t \bigr \} \, d\tau,
\end{equation}
$\alpha $ must be negative in the high-volatility limit \cite{wi}, as assumed in Eq. \ref{sup}.
\vskip 0.25in
\begin{center}    
{\large \bf Long memory}
\end{center}
Volatility has long memory, or is persistent, in the sense that the volatility process decorrelates slowly as the time-lag increases \cite{a1,a2,gh,fo1,ha}. In practice, this means that the decorrelation takes significantly longer than the timescale $\sigma^{-2}$ characterising random fluctuations in the stock price $S$. We then require that the expected percentagechange in volatility over a time step $\Delta t = \sigma^{-2}$ is negligible. From the discretised version of the stochastic differential equation for volatility in Eq. \ref{vol},
\begin{equation}    
\Delta \sigma \approx \alpha \, \Delta t + \beta \,     \bigl [ \, W_2(t+\Delta t) - W_2(t) \, \bigr],     \label{disc}
\end{equation}
we see that expected percentage change in volatility can be approximated as follows,
\begin{equation}    
{{E \{ \Delta \sigma \} } \over {\sigma}} \approx     {\alpha \over \sigma} \, \Delta t.
\end{equation}
The quantity on the LHS is therefore negligible if and only if
\begin{equation}    \vert \alpha \vert \, \sigma^{-3} << 1.
\end{equation}
Equation \ref{pers} arises by requiring this condition to be satisfied in the high--volatility limit.\vskip 0.25in 
\begin{center}    
{\large \bf Volatility is as volatile as the underlying}
\end{center}
Since $\beta/\sigma$ is the volatility of volatility, the meaning of Eq. \ref{vvol} is that volatility remains as volatile as the underlying stock as volatility $\sigma $ tends to infinity. As evidence, Fig. 6.10 in \cite{ta} shows that the volatility of volatility on daily returns typically lies in the range $150\%$ - $450\%$. For comparison, annualised volatility typically lies in the range $5\%$ to $30\%$. Note that we may then expect the limit in Eq. \ref{vvol} to be of order unity or larger. 

\section{Order-of-magnitude estimates}
In this section, we consider a model with only one parameter $r_0$. Using order-of-magnitude estimates of $r_0$ and the characteristic volatility $\sigma_{\max}$ of the most volatile stocks in the market, we argue that the conditions for asymptotic analysis are satisfied for the study conducted by Gopikrishnan {\it et al.}.
\par
Recall that Gopikrishnan {\it et al.} find an inverse cubic decay in a range extending over roughly two decades. Since the return is linear in volatility, the characteristicvolatility $\sigma_{\max} $ of the most volatile stocks in the market will be at least $100 $ times larger than the mean-square volatility, and hence
\begin{equation} \label{est2}  
\sigma^2_{\max} \propto 10000 \, \bigl < \sigma^2 \bigr >,
\end{equation}
where the brackets $\bigl < \cdot \bigr >$ denote time-averaging. (Such volatility levels, however, cannot last for long). Suppose now that the mean-square volatility is $0.04$ per year. Then
\begin{equation}  
\sigma_{\max}^{-2} \propto 1 \, {\rm day}.
\end{equation}
On dimensional grounds, $r_0$ and the mean-square volatility must be proportional 
\begin{equation} \label{est3}    
r_0 \propto \bigl < \sigma^2 \bigr >.
\end{equation}
(This can be proved rigorously by using the expressions for $\alpha $ and $\beta $ in section V to calculate the stationary probability density of volatility). 
With $\Delta t $ equal to $5$ minutes, we get the order-of-magnitude estimates
\begin{eqnarray}     
\sigma_{\max}^2 \, \Delta t & \propto & 0.004 \\     
\sigma_{\max}^2 \, {r_0}^{-1}  & \propto & 10000.
\end{eqnarray}
The first equation states that the largest change in the stock price over a five minutes interval is of the order of $0.4$ percent, while the second equation states that the characteristic diffusiontime-scale is about four orders of magnitude smaller than the characteristic time-scale associated with the interest rate. (The characteristic time-scale associated with the interest rate is the inverse of $r$, and is thus of the order of $30$ years in the EU, for instance). There is then ample empirical support for considering the double limit
\begin{eqnarray} 
\label{small}      
\sigma^2 \, \Delta t & \to & 0 \\      
\sigma^2 \, {r_0}^{-1} & \to & \infty,   \label{li}
\end{eqnarray}
and we now proceed to do so.

\section{Dimensional analysis}
In this section, we use dimensional analysis to establish the general form of 
$\alpha $ and $\beta $. For simplicity, we again consider the special case where there is only one parameter $r_0$. The argument is easily generalised, but at the cost of transparency, to the case where several parameters appear. 
\par
The coefficients $\alpha $ and $\beta $ in the SDE for volatility are then functions in the form,\begin{eqnarray}     
\alpha & = & \alpha(\sigma,r_0) \label{f} \\    
\beta & = & \beta(\sigma, r_0). \label{g}
\end{eqnarray}
Let $[\cdot ]$ denote `dimension' and let $T$ represent time. It is easily shown that
\begin{equation}   
[W] = T^{1/2},
\end{equation}
for the Wiener process $W$, and therefore
\begin{equation}  
[\sigma ] = T^{-1/2}. 
\end{equation}
Using the stochastic differential equation for volatility, we find that
\begin{equation}  
[\alpha]  =  T^{-3/2} 
\end{equation}
and
\begin{equation}
[\beta]  = T^{-1}. 
\end{equation}
The $\Pi$-theorem now implies that
\begin{eqnarray} \label{a1} 
\alpha & = & \sigma^3 \, f(r_0 \, \sigma^{-2}) \\    
\beta & = & \sigma^2 \, g(r_0 \, \sigma^{-2}), \label{a2} 
\end{eqnarray}
for some pair of non-dimensional functions, $f$ and $g$. For $\alpha$ and $\beta $ to be differentiable with respect to $r_0$, as assumed in the introduction, $f(x)$ and $g(x)$ must be differentiable with respect to $x$. 
\section{Asymptotics}
Since $f(x)$ and $g(x)$ are differentiable at zero, 
\begin{eqnarray}  
f(x) & = & f(0) + f'(0) \, x + O(x^2) \\  
g(x) & = & g(0) + g'(0) \, x + O(x^2).
\end{eqnarray}
It then follows from Eqs. \ref{a1} and \ref{a2} that 
\begin{eqnarray} \label{as1}  
\alpha & = & f(0) \, \sigma^{3} + f'(0) \, \sigma \, r_0   + O(r_0^{2} \, \sigma^{-1}) \\  
\beta & = & g(0) \, \sigma^{2} + g'(0) \, r_0 + O(r_0^{2}   \, \sigma^{-2}). \label{as2}
\end{eqnarray}
The only way to satisfy the conditions in Eqs. \ref{sup} --- \ref{vvol} is by taking
\begin{eqnarray}    
f(0) & = & 0 \\    f'(0) & \leq &  0 \\    
g(0) & > & 0.
\end{eqnarray}
Defining
\begin{eqnarray}   
A & = & - f'(0) \\   
B & = & g(0), 
\end{eqnarray}
with 
$A \geq 0$ 
and 
$B>0$, we get 
\begin{equation} \label{model}    
d\sigma = - A \, r_0 \, \sigma \, dt + B \, \sigma^2 dW_2,
\end{equation}
as a model equation for volatility in the limit
\begin{equation} \label{lim}   
\sigma^{2} \, r_0^{-1} \to \infty.
\end{equation} 
The solution of Eq. \ref{model} is superdiffusive in the high-volatility limit, i.e. it diffuses much faster than Brownian motion. The fastest increase in $\sigma$ occurs for $A=0$. Even then, the solution remains finite almost surely at all finite times (p. 332, \cite{ka}).  

\section{The probability density of volatility}
In this section, we consider the tails of the stationary probability density $q(\sigma)$ of volatility. Stationarity is a reasonable approximation when sampling frequently over long intervals. Recall that the authors of Ref. \cite{go1} sample returns at $5$ to $120$ minutes intervals over a two--year period. Since $q(\sigma)$ satisfies the stationary version of the Kolmogorov forward equation \cite{ka},
\begin{equation} \label{kolm}   
{{\partial} \over {\partial \sigma}} \Bigl [ \, \alpha \,    q \,    \Bigr ] =    {1 \over 2} \, {{\partial^2} \over {\partial \sigma^2} }    \Bigl [ \, \beta^2 \, q \, \Bigr ],
\end{equation}
it follows from Eq. \ref{model} that
\begin{equation} \label{s2}   
q(\sigma) \sim C_0 \,  r_0^{3/2} \, \sigma^{-4},
\end{equation}
as $\sigma^2 \, r_0^{-1}$ tends to infinity, where $C_0$ is a non-dimensional constant. 
\par
After the present paper was first submitted for publication, Ref. \cite{go2} reported empirical evidence for Eq. \ref{s2} .

\section{The tail probability}
We now calculate the tail probability $\bar P(x)$ in the short--term high--volatility limit. Specifically, we assume 
\begin{eqnarray*}  
\sigma^2/r_0 & \to & \infty \\  
\sigma^2 \, \Delta t & \to & 0.
\end{eqnarray*}
The approximation in Eq. \ref{ap} is valid in this limit,
\begin{equation}    
X =_d \sigma \, \bigl [ \, W_1(t+\Delta t) - W_1(t)        \, \bigr ] + O(\Delta t).
\end{equation}
so that $X$ can be approximated, in distribution, by the product of two independent random variables. Using a well known result for the probability density of a product of two independent random variables (e.g., Sect. 4.7, \cite{gi}), we may express the probability density $p(x)$ of $X$as follows,
\begin{equation} \label{prod}   
p(x) =  {1 \over {\sqrt {2 \, \pi \, \Delta t}} }  \int_{-\infty}^{+\infty} q(z)  \, 
              \exp \bigl [ -  x^2/2 \, z^{2}  {\Delta t} \, \bigr ] \,  \vert z \vert^{-1}   \, dz + O(\Delta t).
\end{equation}
After inserting the expression for $q$ inEq. \ref{s2} and integrating, we obtain 
\begin{equation} \label{heureka}   
\bar P(x) \sim C \, r_0^{3/2} \, (\Delta t)^{3/2} \, x^{-3},
\end{equation}
where
\begin{equation}    
C = C_0/3 \, \sqrt{2/{ \pi}}   \, \int_{-\infty}^{+\infty} \, \vert z \vert^{-5}  \, \exp \bigl [ - z^{-2}/2 \,               \bigr ] \, dz.
\end{equation}
\par
Thus, when choosing a stochastic volatility model that reproducesstylised facts about volatility, we find that the tail probability of the logarithmic return falls off like an inverse cubic, as reported from empirical evidence in Ref. \cite{go1}. The prediction that the tail probability $\bar P(x)$ scales like $(\Delta t)^{3/2}$ seems to be new. \vskip 0.5in{\bf Acknowledgements} 
\par \noindent
Henrik O. Rasmussen would like to acknowledge financial support from the European Union, in the form of a TMR Fellowship.

\begin{thebibliography}{99}
\bibitem{ma}  B. B. Mandelbrot, The variation of certain speculative              prices, {\it Journal of               Business}, {\bf 36}, 394 -- 419 (1963).
\bibitem{fa}  E. F. Fama, Mandelbrot and the stable Paretian hypothesis,               {\it Journal of Business},              {\bf 36}, 420 -- 429, (1963).
\bibitem{man} R. N. Mantegna \& H. E. Stanley,               Scaling behaviour in the dynamics of an economics               index, {\it Nature}, {\bf 376}, 46 -- 49, (1995). 
\bibitem{eb}  E. Eberlein \& U. Keller, Hyperbolic distributions in              finance, {\it Bernoulli},              {\bf 1}, 281 -- 299, (1995).
\bibitem{go1}  P. Gopikrishnan, M. Meyer, L. A. N. Amaral, and H. E. Stanley,               {Inverse cubic law for the distribution of stock price variations},               {\it Eur. Phys. J.} B {\bf 3}, 139 -- 140 (1998).
\bibitem{hull} J. Hull, {\it Options, Futures, \& Other Derivatives},                (Prentice--Hall, 2000).\bibitem{bl}  F. Black \& M. Scholes, The pricing of options and corporate              liabilities, {\it Journal of Political Economy}, {\bf 81},               637 -- 654, (1973).  
\bibitem{me}  R. C. Merton, {\it Continuous--time finance}, (Blackwell, 1990).
\bibitem{hu}  J. Hull \& A. White, The pricing of options on assets              with stochastic volatilities,              {\it J. Fin.} {\bf 42} 281 -- 300 (1987). 
\bibitem{du}  D. Duffie, {\it Dynamic Asset Pricing Theory}, Princeton University              Press (1992).
\bibitem{wi}  P. Wilmott, {\it Derivatives}, (Wiley, 1998). 
\bibitem{bou} J-P. Bouchaud, D. Sornette, C. Walter, \& J. P. Aguilar,              Taming large events: optimal portfolio theory for              strongly fluctuating assets, {\it              Int. J. Theor. Appl. Fin.} {\bf 1}, 25 -- 42, (1999).   
\bibitem{bi}  G. Birkhoff, {\it Hydrodynamics, A Study in Logic, Fact, and Similitude},              Princeton University Press (1950). 
\bibitem{ba}  G. I. Barenblatt, {\it Similarity, Self-Similarity, and Intermediate               Asymptotics}, (English translation 1979, Consultants Bureau).
\bibitem{to}  R. G. Tompkins, {\it Option Analysis}, Irwin (1994). 
\bibitem{a1}  T. G. Andersen \& T. Bollerslev,               {Heterogeneous information arrivals and return volatility              dynamics:               uncovering the long-run in high frequency returns},              {\it J. Finance}, {\bf LII} no. 3, 975 -- 1005 (1997).
\bibitem{a2}  T. G. Andersen \& T. Bollerslev,               {Deutsche Mark - Dollar volatility: intraday activity              patterns, macroeconomic announcements,              and longer run dependencies},              {\it J. Finance}, {\bf LIII} no. 1, 219 -- 265 (1998).
\bibitem{gh}  C. A. E. Goodhart \& M. O'Hara,               {High frequency data in financial markets: issues and applications},              {\it Journal Empirical Finan.}, {\bf 4}, 73 -- 114, (1997).
\bibitem{fo1} J.-P. Fouque, G. Papanicolaou, \& K. R. Sircar,              Financial modeling in a fast mean--reverting stochastic volatility environment,              {\it Asia--Pacific Financial Markets}, {\bf 6}, 37 -- 48, (1999).
\bibitem{ha} A. Harvey, Long memory in stochastic volatility,             In {\it Forecasting             volatility in the financial markets}, Eds. J. Knight \& S.             Satchell, (Heinemann, 1998).\bibitem{ta} N. Taleb, {\it Dynamic Hedging: Managing Vanilla and Exotic Options},              John Wiley and Sons (1997).
\bibitem{ka} I. Karatzas \& S. Shreve, {\it Brownian Motion and             Stochastic Calculus} (2nd ed.),  Springer Verlag (1991).
\bibitem{go2}              Y. Liu, P. Gopikrishnan, P. Cizeau, M. Meyer, C.-K. Peng, and H. E. Stanley,              {Statistical properties of the volatility of price fluctuations},              {\it Phys. Rev. E} {\bf 60}, 1390 -- 1400 (1999).
\bibitem{gi} G. R. Grimmett \& D. R. Stirzaker,              {\it Probability and Random Processes} (2nd ed.),             Oxford University Press (1992).
\end{thebibliography}
\end{document}